\documentclass[preprint,12pt]{revtex4}

\usepackage{graphicx}
\usepackage{amssymb}
\usepackage{amsmath}
\usepackage{subfigure}

\begin{document}

\title{A Stable and Robust Calibration Scheme \\of the Log-Periodic Power Law Model}

\author{Vladimir Filimonov $\dag \ddag$ and Didier Sornette $\dag \S$}
\email{vfilimonov@ethz.ch, dsornette@ethz.ch}
\thanks{\\~Corresponding author: Didier Sornette.}
\address{\normalfont{$\dag$ 
Department of Management, Technology and Economics\\ ETH Z\"{u}rich\vspace{12pt}\\ 
$\ddag$ Laboratory of quantitative analysis and modeling of economics\\
National Research University --- Higher School of Economics, Nizhny Novgorod, Russia \vspace{12pt}\\ 
$\S$ Swiss Finance Institute, c/o University of Geneva}\\\vspace{12pt}}

\begin{abstract}
We present a simple transformation of the formulation of the log-periodic power law
formula of the Johansen-Ledoit-Sornette model of financial bubbles that reduces
it to a function of only three nonlinear parameters. 
The transformation significantly decreases the complexity of the fitting procedure and 
improves its stability tremendously because the modified cost function is now characterized by
good smooth properties with in general a single minimum in the case where the model
is appropriate to the empirical data. We complement the approach with 
an additional subordination procedure that slaves two of the nonlinear parameters
to what can be considered to be the most crucial nonlinear parameter, the critical time $t_c$ defined
as the end of the bubble and the most probably time for a crash to occur. 
This further decreases the complexity of the search and provides an intuitive
representation of the results of the calibration. With our proposed methodology, metaheuristic searches are not longer
necessary and one can resort solely to rigorous controlled local search algorithms, leading
to dramatic increase in efficiency. Empirical tests on the 
Shanghai Composite index (SSE)  from January 2007 to March 2008 illustrate our findings.

\vskip 1cm
{\bf Keywords:} JLS model, financial bubbles, crashes, log-periodic
power law, fit method, optimization.

\end{abstract}
\keywords{}

\maketitle \clearpage

\section{Introduction}
\label{sec:intro}

Financial crises are crippling national economies, as evidenced by the episode of the ``great depression'' that
followed the great crash of October 1929 and by the ``great recession'' that followed the financial debacle that
started in 2007 with the cascade of defaults of financial debt securities. In these two cases \cite{Galbraith1997,Sorwoodtokyo}, 
as well as in many others \cite{Sornette_Crash2002},  one can observe that the crisis followed the burst of one or several bubbles, defined 
qualitatively as an exaggerated leverage in some industry sector. 
This observation has led policy officials to call for the development of methodologies
aiming at diagnosing the formation of bubbles as early as possible in order to take
appropriate counter measures.

The problem is however extremely difficult, because the definition of what is a bubble is prone
to controversies. Superficially, financial bubbles are easily defined as transient upward accelerations of 
the observed price above a fundamental value~\cite{Galbraith1997,Kindleberger2000,Sornette_Crash2002}. 
The paradox is that the determination of a bubble requires, in this definition, a precise
determination of what is the fundamental value. But, the fundamental value is in general poorly 
constrained. In addition, a transient exponential acceleration of the observed price that would be taken
as the diagnostic of a developing bubble is not 
distinguishable from an exponentially growing fundamental price. 

The Log Periodic Power Law (LPPL) proposed by A. Johansen, O. Ledoit and D. Sornette (JLS)~\cite{SornetteJohansen1999Risk,SornetteJohansen2000,SornetteJohansen1999,Sornette_Crash2002} 
proposes a way out of this dead-end by defining a bubble as a transient  ``faster-than-exponential'' growth,
resulting from positive feedbacks. The JLS model
provides a flexible framework to detect bubbles and predict changes of regime from
the study of the price time series of a financial asset.  In contrast to 
the more traditional view, the bubble is here defined as a faster-than-exponential increase in asset prices, that reflects positive feedback loop of higher return anticipations competing with negative feedback spirals of crash expectations. The Johansen-Ledoit-Sornette 
models a bubble price as a power law with a finite-time singularity decorated by oscillations with a  frequency
increasing with time.   

The LPPL model in its original form presents a function of 3 linear and 4 nonlinear parameters that should be estimated by fitting this function to the log-price time series. Calibrating the LPPL model has always been prone with difficulties due to the relatively
large number of parameters that must be estimated and the strong nonlinear structure of the equation.
Most of the fitting procedures that have been used until now
subordinate the 3 linear parameters to the 4 nonlinear parameters~\cite{SornetteJohansen2000}.
The resulting search space with 4 nonlinear parameters has a very complex quasi-periodic structure with multiple minima.
The determination of the global minimum requires using some metaheuristic methods such as taboo search~\cite{SornetteJohansen2000,SornetteJohansen2001} or genetic algorithm~\cite{Jacobsson2009}. But even these methods
do not ensure that the correct solution is discovered. In addition, how to deal with the existence of many possible competing
degenerate solutions has not been satisfactorily solved. 

In the present paper, we propose a fundamental revision of the formulation of the LPPL model, that transforms
it from a function of 3 linear and 4 nonlinear parameters into a representation with 4 linear and 3 nonlinear parameters.
This transformation significantly decreases the complexity of the fitting procedure and 
improves its stability tremendously because the modified cost function is now characterized by
good smooth properties with in general a single minimum in the case where the model
is appropriate to the empirical data. 
As an additional step, we propose a subordination procedure that slaves two of the nonlinear parameters
to what can be considered to be the most crucial nonlinear parameter, the critical time $t_c$ defined
as the end of the bubble. The critical time is indeed the prize of the whole forecasting exercise: the 
sooner a bubble is identified, that is, the further away is $t_c$ from the present time, the better it is 
for policy makers to take appropriate actions. Of course, it goes without saying that the validity and
reliability of the results should be established carefully before any action is taken. We note that this
additional subordination decreases further the complexity of the search and provides an intuitive
representation of the results of the calibration. With our proposed methodology, metaheuristics are not longer
necessary and one can resort solely to rigorous controlled local search algorithms, leading
to dramatic increase in efficiency.

The paper is organized as follows. Section 2 describe the idea behind the LPPL model and presents its original form. Traditional fitting procedures are described in section 3. Section 4 presents the proposed modification of the model and
explain the new subordination procedure. Section 5 concludes.

\section{Log-Periodic Power Law model}
\label{sec:lppl}

The JLS model~\cite{SornetteJohansen1999Risk,SornetteJohansen2000,SornetteJohansen1999} assumes 
that the logarithm of the asset price $p(t)$ follows a standard diffusive dynamics with 
varying drift $ \mu(t)$ in the presence of discrete discontinuous jumps:
\begin{equation}\label{eq:JLS}
    \frac {dp}{p} = \mu(t)dt+\sigma(t)dW-\kappa dj~.
\end{equation}
In this expression, $\sigma(t)$ is the volatility, $dW$ is the infinitesimal 
increment of a standard Wiener process and $dj$ represents a discontinuous jump such as $j=0$ before 
the crash and $j=1$ after the crash occurs ($\int_{t_{\rm before}}^{t_{\rm after}} dj =1$). 
The parameter $\kappa$ quantifies the amplitude of the crash when it occurs. The expected value of $dj$ 
is nothing but the crash hazard rate $h(t)$ times in the infinitesimal time increment $dt$:  $\text{E}[dj]=h(t)dt$. 
The JLS model assumes that two types of agents are present in the market: a group of traders 
with rational expectations and a group of noise traders who exhibit herding behavior 
that may destabilize the asset price. According to the JLS model,
the actions of noise traders are quantified by the following dynamics of the hazard rate~\cite{SornetteJohansen1999Risk,SornetteJohansen2000,SornetteJohansen1999}:
\begin{equation}\label{eq:hazard}
    h(t)=\alpha(t_c-t)^{m-1}\big(1+\beta\cos(\omega\ln(t_c-t)-\phi')\big),
\end{equation}
where $\alpha$, $\beta$, $\omega$ and $\phi$ are parameters and $t_c$ is the critical time 
that corresponds to the end of the bubble. The power law behavior $(t_c-t)^{m-1}$ embodies 
the mechanisms of positive feedback at the origin of the formation of bubble. 
The log-periodic function $\cos(\omega\ln(t_c-t)-\phi')$ takes into account the existence of a possible hierarchical 
cascade of panic acceleration punctuating the course of the bubble. 
The no-arbitrage condition $\text{E}[dp]=0$ imposes that the excess return $\mu(t)$ is proportional
to the crash hazard rate $h(t)$: $\mu(t) = \kappa h(t)$. Solving 
equation~\eqref{eq:JLS} with $\mu(t) = \kappa h(t)$ and under the condition that no crash has yet occurred ($dj=0$)
leads to the following log-periodic power law (LPPL) equation for the expected value of a log-price:
\begin{equation}\label{eq:lppl}
   \text{E}[ \ln p(t)]=A+B(t_c-t)^m+C(t_c-t)^m\cos(\omega\ln(t_c-t)-\phi),
\end{equation}
where $B=-\kappa\alpha/m$ and $C=-\kappa\alpha\beta/\sqrt{m^2+\omega^2}$. 
It should be noted that solution~\eqref{eq:lppl} describes the dynamics of the average log-price only up to critical time $t_c$ and 
cannot be used beyond it. This critical time $t_c$ corresponds to the termination of the bubble and indicates the change to another regime, which could be a large crash or a change of the average growth rate.

The LPPL model~\eqref{eq:lppl} is described by 3 linear parameters ($A,B,C$) 
and 4 nonlinear parameters ($m,\omega,t_c,\phi$). These parameters 
are subjected to the following constrains. Since the integral of the hazard rate~\eqref{eq:hazard} over time up to $t=t_c$ gives the probability of 
the occurrence of a crash, it should be bounded by $1$, which yields the condition $m<1$. At the same time, 
the log-price~\eqref{eq:lppl} should also remain finite for any $t\leq t_c$, which imply the other condition $m>0$. 
In addition, the requirement of the existence of an acceleration of the hazard rate 
as time converges towards $t_c$ implies $B<0$. Additional constraints emerge from a compilation of 
a significant number of historical bubbles \cite{SornetteJohansen2001,SornetteJohansen2010,SornetteLin2009}
that can be summarized as follows:
\begin{equation} \label{eq:constrains}
    0.1\leq m\leq0.9,\quad 6\leq\omega\leq13,\quad |C|<1,\quad B<0.
\end{equation}
These conditions~\eqref{eq:constrains} can be regarded as the ``stylized features of LPPL''. 
The condition $6\leq\omega\leq13$ constrains the log-periodic oscillations to be neither too fast (otherwise they would fit the random component of the data), nor too slow (otherwise they would provide a contribution to the trend 
(see Ref.~\cite{SornetteJohansen2000_geophisics} in this respect for the conditions on the statistical significance of log-periodicity). The last restriction $|C|<1$ in~\eqref{eq:constrains} was introduced in~\cite{Bothmer2003} to ensure that the hazard rate $h(t)$ remains always positive.

\section{Summary of existing fitting procedure of the LPPL function (\ref{eq:lppl})}
\label{sec:fitting}

Forecasting the termination of a bubble
amounts to finding the best estimation of the critical time $t_c$. This requires calibrating the LPPL formula (\ref{eq:lppl})
to determine the parameters $t_c,m,\omega,\phi,A,B,C$ of the model that best fit some observed price time series $p(t)$ within
a time window $t\in[t_1,t_2]$. This may be performed using the Least-Squares Method of minimizing the sum of squared residuals:
\begin{equation}\label{eq:S}
  S(t_c,m,\omega,\phi,A,B,C)=\sum_{i=1}^{N} \Big[
  \ln p(\tau_i) - A - B (t_c - \tau_i)^m - C  (t_c - \tau_i)^m \cos(\omega \ln(t_c -\tau_i) - \phi)
  \Big]^2,
\end{equation}
where $\tau_1=t_1$ and $\tau_N=t_2$.
Minimization of such nonlinear multivariate cost function $S$ is a non-trivial task due to presence of multiple local minima, where the local optimization algorithm can get trapped. However, the complexity of the optimization problem may be significantly decreased by noticing that three linear parameters $A,B,C$ can be slaved to the four other nonlinear parameters $t_c,m,\omega,\phi$ \cite{SornetteJohansen2000}. 
Indeed, it is easy to prove that 
\begin{equation}\label{eq:min_1}
  \min_{t_c,m,\omega,\phi,A,B,C}S(t_c,m,\omega,\phi,A,B,C)\equiv
  \min_{t_c,m,\omega,\phi}S_1(t_c,m,\omega,\phi), 
\end{equation}
where
\begin{equation}\label{eq:S1}
  S_1(t_c,m,\omega,\phi)=
  \min_{A,B,C}S(t_c,m,\omega,\phi,A,B,C).
\end{equation}
The last optimization problem~\eqref{eq:S1} may be rewritten as:
\begin{equation}\label{eq:S1_arg}
  \{\hat A, \hat B, \hat C\}=\arg\min_{A,B,C}S(t_c,m,\omega,\phi,A,B,C)=
  \arg\min_{A,B,C}\sum_{i=1}^{N} \Big[ y_i - A - B f_i - C  g_i\Big]^2,
\end{equation}
where $y_i=\ln p(\tau_i)$, $f_i=(t_c-\tau_i)^m$, $ g_i=(t_c-\tau_i)^m\cos(\omega\ln(t_c-\tau_i)-\phi)$. Being linear 
in terms of the variables $A,B,C$, for $m\ne0$, $\omega\ne0$, $t_c>t_2$ and for any $\phi$, this problem has one unique solution 
that an explicit analytical solution obtained from the first order condition. The first order condition leads to the matrix equation
\begin{equation}\label{eq:ABC}
  \left(\begin{array}{ccc}
    N & \sum f_i & \sum g_i \\
    \sum f_i & \sum f_i^2 & \sum f_ig_i \\
    \sum g_i & \sum f_ig_i & \sum g_i^2 \\
  \end{array}\right)\left(\begin{array}{c}
    \hat A\\ \hat B\\ \hat C  
  \end{array}\right)=
  \left(\begin{array}{c} 
    \sum y_i \\ \sum y_if_i \\ \sum y_ig_i \\
  \end{array}\right)
\end{equation}
which is solved in a standard way using the LU decomposition 
algorithm~\cite{Press2007NumericalRecipes}. Then, the global fitting procedure is reduced to solving the nonlinear optimization problem
\begin{equation}\label{eq:S_arg}
  \{\hat t_c,\hat m,\hat \omega,\hat \phi\}=\arg\min_{t_c,m,\omega,\phi}S_1(t_c,m,\omega,\phi),
\end{equation}
where the cost function is given by expression \eqref{eq:S1}.

\begin{figure}[htbp]
  \centering
  \includegraphics[width=0.7\textwidth]{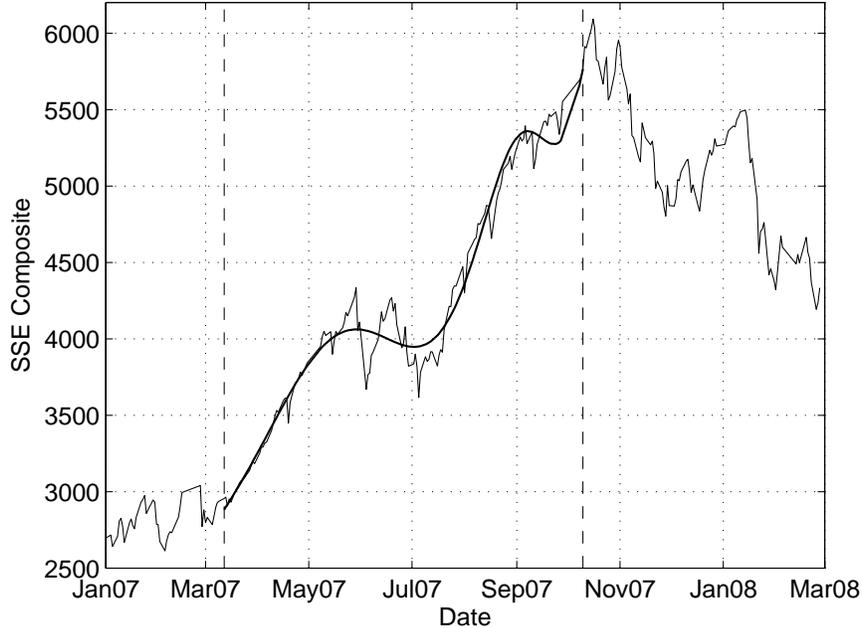}
  \caption{Time dependence of the SSE Composite Index (Shanghai Composite index) from January 2007 to January 2008 (thin noisy solid line) and the best fit obtained by the LPPL model~\eqref{eq:lppl} (thick solid line). Vertical dashed lines delineate the time window $[t_1,t_2]$ used
  in the fitting procedure.} \label{fig:ssec}
\end{figure}

\begin{figure}[htbp]
  \centering
   \subfigure[$S_1(t_c, m=0.7, \omega=7.5, \phi)$]
   { \includegraphics[width=0.48\textwidth]{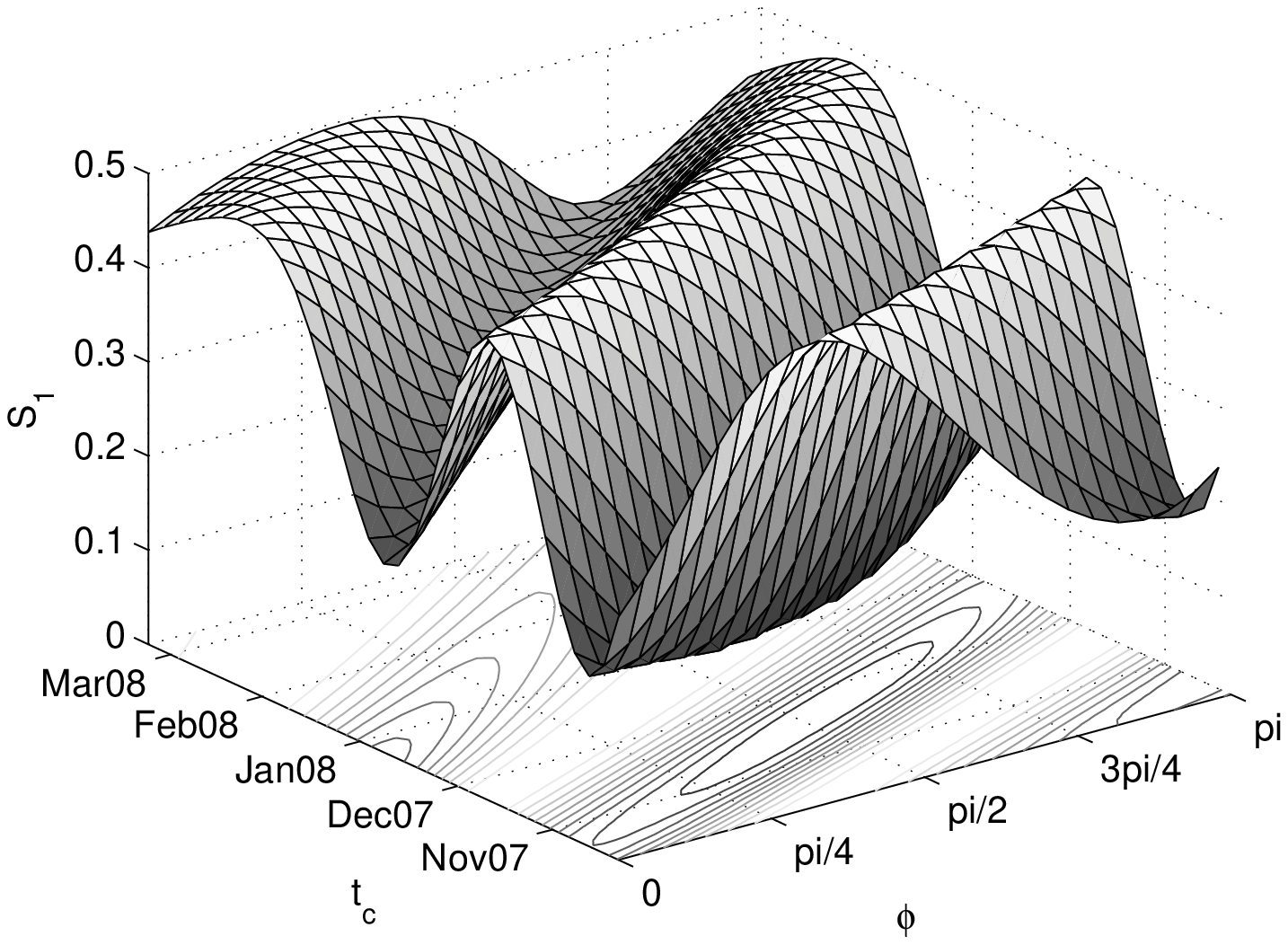}} 
   \subfigure[$S_1(t_c=\mbox{'06-Nov-2007'}, m, \omega, \phi=\pi/10)$]
   { \includegraphics[width=0.48\textwidth]{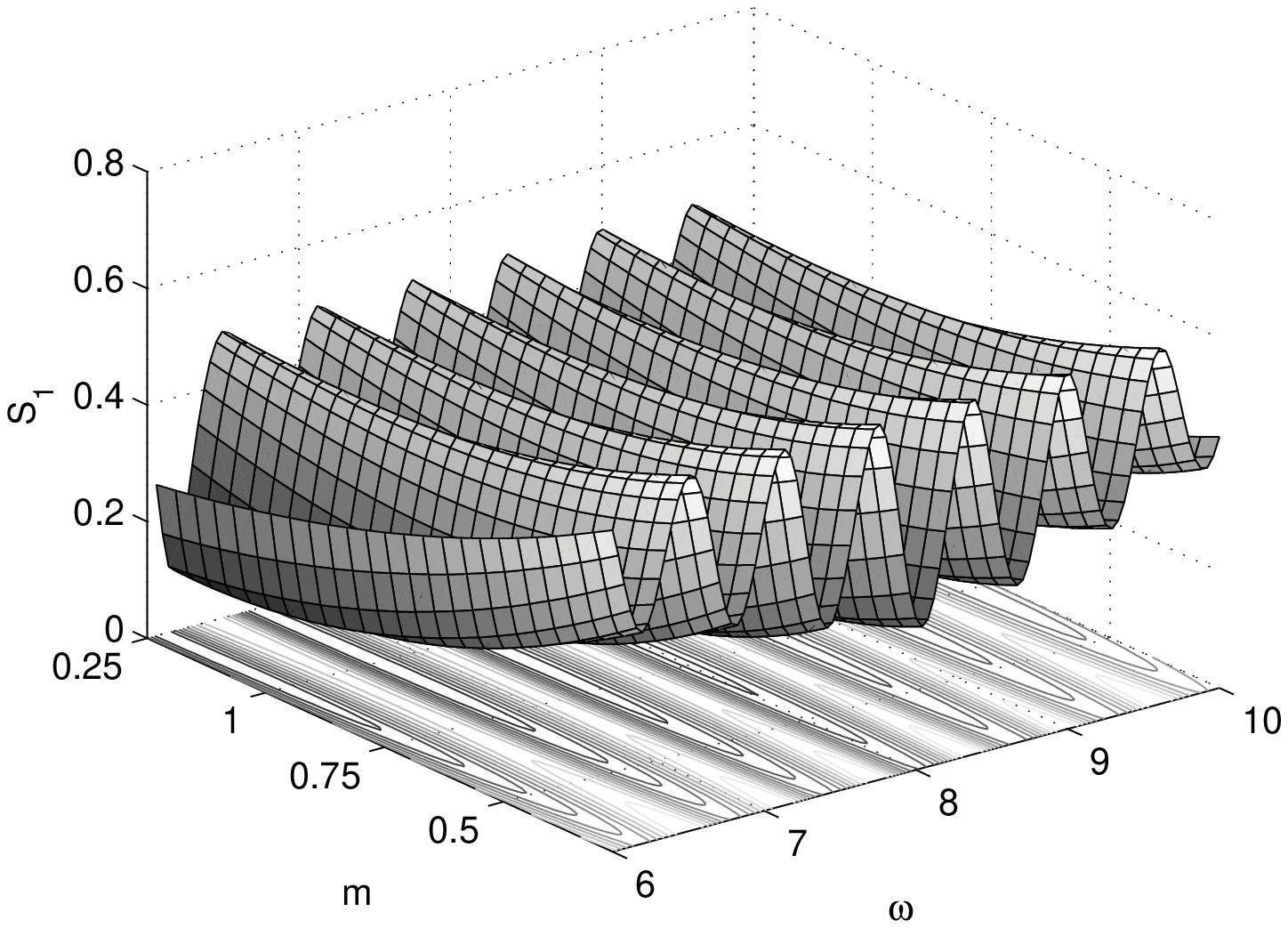}}\\
   \subfigure[$S_1(t_c, m=0.7, \omega, \phi=\pi/10)$]
   { \includegraphics[width=0.48\textwidth]{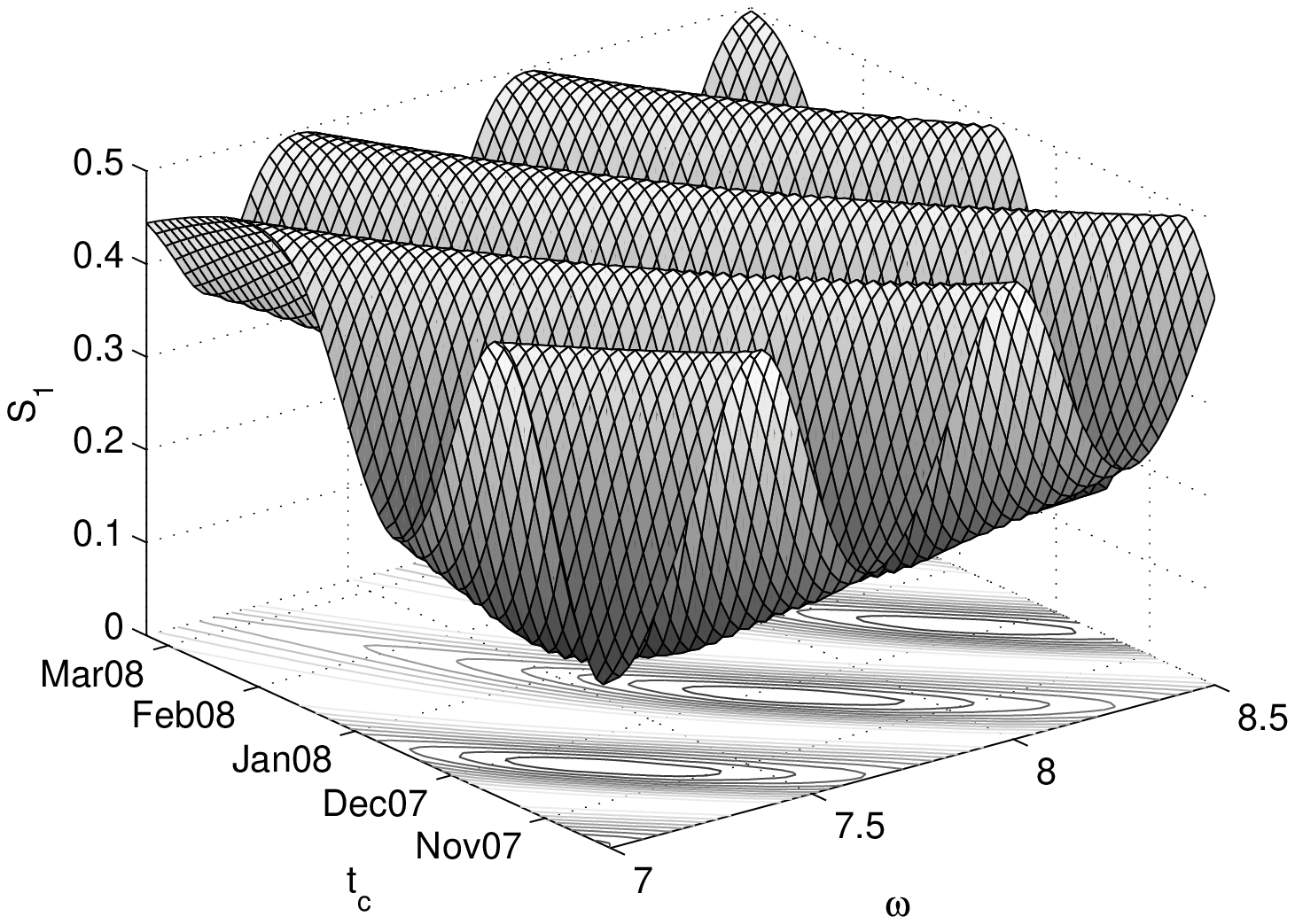}}
   \subfigure[$S_1(t_c=\mbox{'06-Nov-2007'}, m, \omega=7.5, \phi)$]
   { \includegraphics[width=0.48\textwidth]{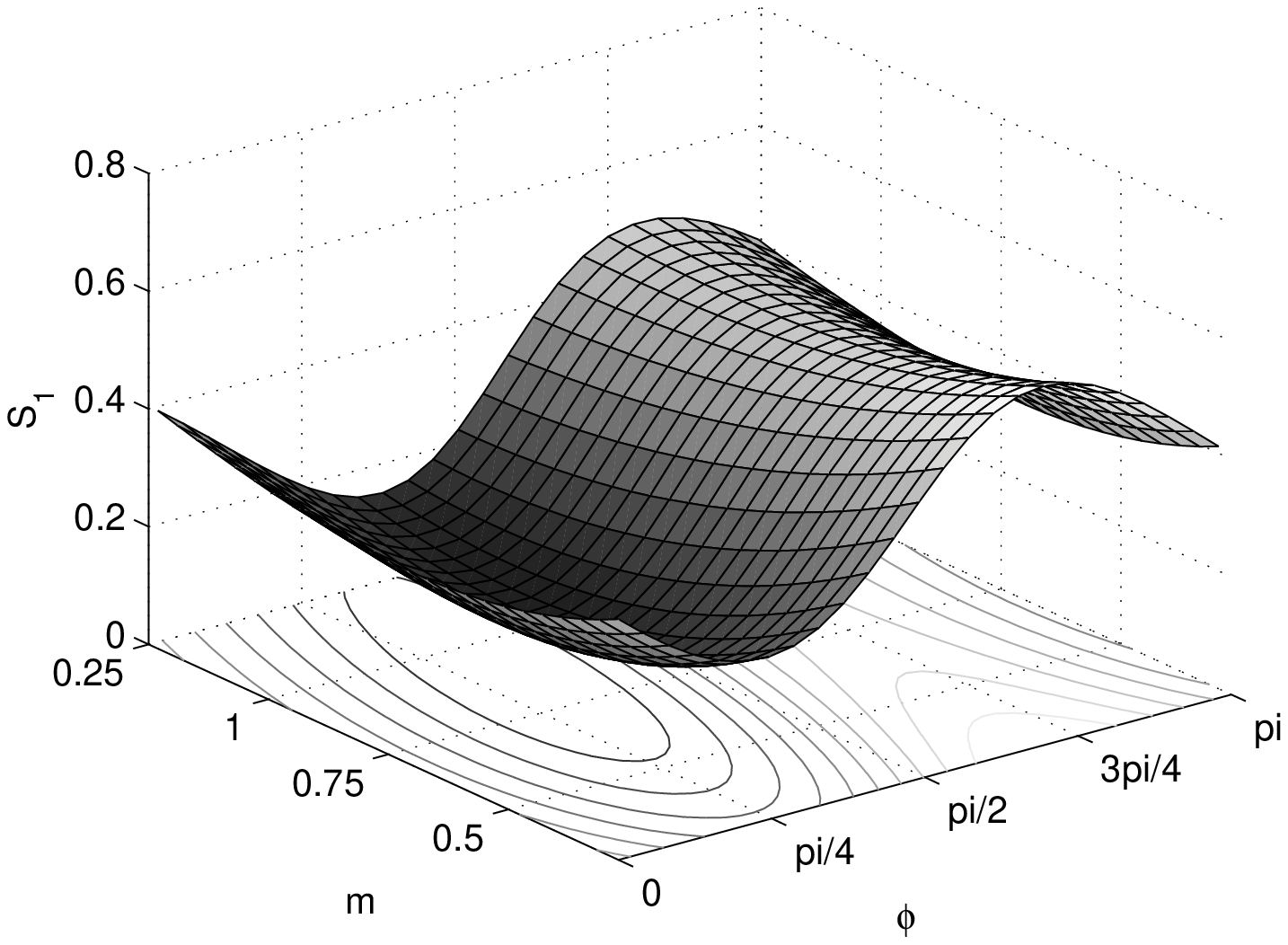}}
   \caption{Different cross-sections of the cost function $S_1$~\eqref{eq:S1} for the time-series of SSE Composite Index presented at fig.~\ref{fig:ssec}.} \label{fig:crosssections}
\end{figure}

Reducing the number of parameters from 7 to 4 simplifies considerably the calibration problem.
However, the minimization problem~\eqref{eq:S_arg} still requires finding the global minimum in 
a four dimensional space of a cost function~\eqref{eq:S1} with multiple extrema. The complexity of a
typical cost function is illustrated by the different cross-sections shown in fig.~\ref{fig:crosssections}.
This figure is obtained for the daily time-series of the Shanghai Composite index (SSE)  from January 2007 to March 2008, shown in fig.~\ref{fig:ssec}. Here, for fitting purposes, the time window between $t_1=\mbox{'12-Mar-2007'}$ and $t_2=\mbox{'10-Oct-2007'}$ was considered. 

The complex multiple extrema structure of the cost function does not allow one to use  local search algorithms, such as 
the steepest descent or the Newton's method, in order to find the solution of~\eqref{eq:S_arg}. Such complex optimization problems were studied extensively in the last 30 years and a number of approaches developed mostly within the class of so-called metaheuristic algorithms~\cite{Talbi2009Metaheuristics} were introduced to solve this class of problems. Metaheuristic algorithms usually make only a few or no assumptions on the cost function, do not require calculating the derivatives and can search a very large multidimensional spaces of candidate solutions. The cost of such universality and generality is the absence of any guarantee of finding the optimum or even a satisfactory near-optimal solution. This issue is usually solved with the combination of an initial metaheuristic explorative search followed by a local descent to the closest minimum in a second step. 

For the problem of calibrating the LPPL function to financial price time series, 
the taboo search \cite{Cvijovicacute1995} has been the main approach 
\cite{SornetteJohansen2000,SornetteJohansen2001,SornetteZhouIJF06}. 
Taboo search enhances the performance of a local search method by using memory structures that describe the visited solutions: once a potential solution has been determined, it is marked as ``taboo'' area, so that the algorithm does not visit explored regions repeatedly. 
Being metaheuristic, the taboo search does not guarantee convergence. Thus in the original work~\cite{SornetteJohansen1999Risk},
it was proposed to keep the 10 best outcomes of the taboo search as initial conditions 
for a local Levenberg-Marquardt nonlinear least squares algorithm~\cite{Levenberg1944,Marquardt1969}. The solution with the minimum sum of squares between the fitted model and the observations is then taken as the final solution. 

Another important metaheuristic algorithm that was proposed and successfully used to fit the LPPL function 
to financial time series is the genetic algorithm~\cite{Jacobsson2009}. The genetic algorithm 
mimicks the natural selection process occurring in biological systems, and is governed by four phases: a selection mechanism, a breeding mechanism, a mutation mechanism and a culling mechanism~\cite{Gulsen1995}. Similar to the taboo search, the results of the genetic algorithm
were refined in a second step by using them as a starting values for the Nelder-Mead simplex method~\cite{NelderMead1965}.

Using any of the fitting procedures above with expression \eqref{eq:lppl} always yield some result. However, this does
not mean that this result should be trusted. The constraints \eqref{eq:constrains} constitute a convenient approach that
has proven very useful to
filter the solutions that are believed to meaningful and
relevant to describe developing bubbles from the spurious ones (those that correspond to ``fitting an elephant''
with formulas that contain 4 or more parameters. Recall Enrico Fermi as cited by Freeman Dyson
who said: ``I remember my friend Johnny von Neumann used to say, with four parameters I can fit an
elephant, and with five I can make him wiggle his trunk.").

As already mentioned, the procedures based on the taboo search and 
on genetic algorithms, even when supplemented with the local search algorithms, do not guarantee convergence to the best solution of
the optimization problem~\eqref{eq:S_arg}. This opens the gates for criticisms directed to any of such fitting procedures
\cite{Rosser-2008-ACS,Lux-2009,BreeJoseph2010, BreeChallet2010}. In the present paper, 
 we present a reformulation of the optimization problem~\eqref{eq:S_arg} that simplifies considerably the fitting method.
 Applied to previous calibrations using the procedures described in this section 
 \cite{SornetteJohansen1999,SornetteJohansen1999Risk,SornetteJohansen2000,SornetteJohansen2001,SornetteZhouIJF06,SornetteJohansen2010,SornetteZhou2010}, we are able to confirm the essential goodness of fits of the obtained calibrations on empirical
 financial time series.

\section{New fitting method for the log-periodic power law}
\label{sec:newfitting}

This section presents the new calibration method of the LPPL model, which can be formulated in two steps
that are successively described in the coming two subsections.

\subsection{From 3 to 4 slaved linear parameters by transformation of the phase}

The key ideas of the new proposed methods is to decrease the 
number of nonlinear parameters and get rid at the same time
of the interdependence between the phase $\phi$ and the angular log-frequency $\omega$.
For this, we rewrite the LPPL formula \eqref{eq:lppl} by expanding the cosine term as follows:
\begin{equation}\label{eq:lppl2}
  \ln\text{E}[ p(t)] = A + B(t_c - t)^m + C(t_c - t)^m \cos(\omega \ln(t_c -t))\cos\phi + C(t_c - t)^m \sin(\omega \ln(t_c -t))\sin\phi. 
\end{equation}
Let us introduce two new parameters 
\begin{equation}\label{eq:C12}
  C_1=C\cos\phi,	\quad 	C_2=C\sin\phi
\end{equation}
and rewrite the LPPL equation~\eqref{eq:lppl} as:
\begin{equation}\label{eq:lppl_new}
  \ln\text{E}[ p(t)] = A + B(t_c - t)^m + C_1(t_c - t)^m \cos(\omega \ln(t_c -t)) + C_2(t_c - t)^m \sin(\omega \ln(t_c -t)). 
\end{equation}
As seen from~\eqref{eq:lppl_new}, the LPPL function has now only 3 nonlinear ($t_c, \omega, m$) and 4 linear $A,B,C_1,C_2$ parameters,
and the two new parameters $C_1$ and $C_2$ contain formerly the phase $\phi$. 

As in the previous section, in order to estimate the parameters, we use the least-squares method with cost function
\begin{multline}\label{eq:S_new}
  F(t_c,m,\omega,A,B,C_1,C_2)=
  \sum_{i=1}^{N} \Big[
  \ln p(\tau_i) - A - B(t_c - \tau_i)^m - \Big.\\\Big.
   C_1(t_c - \tau_i)^m \cos(\omega \ln(t_c -\tau_i)) - C_2(t_c - \tau_i)^m \sin(\omega \ln(t_c -\tau_i))
  \Big]^2.
\end{multline}
Slaving the 4 linear parameters $A,B, C_1,C_2$ to the 3 nonlinear $t_c, \omega, m$, we obtain the nonlinear optimization problem
\begin{equation}\label{eq:S_arg_new}
  \{\hat t_c,\hat m,\hat \omega\}=\arg\min_{t_c,m,\omega}F_1(t_c,m,\omega),
\end{equation}
where the cost function $F_1(t_c,m,\omega)$ is given by
\begin{equation}\label{eq:S1_new}
  F_1(t_c,m,\omega)=
  \min_{A,B,C_1, C_2}F(t_c,m,\omega,A,B,C_1, C_2).
\end{equation}
Similarly to the procedure \eqref{eq:S1_arg} leading to \eqref{eq:ABC}, the optimization problem ($\{\hat A, \hat B, \hat C_1, \hat C_2\}=\arg\min_{A,B,C_1,C_2}F(t_c,m,\omega,A,B,C_1,C_2)$) has a unique solution obtained from the matrix equation:
\begin{equation}\label{eq:ABC}
  \left(\begin{array}{cccc}
    N & \sum f_i & \sum g_i & \sum h_i \\
    \sum f_i & \sum f_i^2 & \sum f_ig_i & \sum f_ih_i\\
    \sum g_i & \sum f_ig_i & \sum g_i^2  & \sum g_ih_i\\
    \sum h_i & \sum f_ih_i & \sum g_ih_i  & \sum h_i^2\\
  \end{array}\right)\left(\begin{array}{c}
    \hat A\\ \hat B\\ \hat C_1 \\ \hat C_2  
  \end{array}\right)=
  \left(\begin{array}{c} 
    \sum y_i \\ \sum y_if_i \\ \sum y_ig_i \\ \sum y_ih_i
  \end{array}\right)
\end{equation}
where $y_i=\ln p(\tau_i)$, $f_i=(t_c-\tau_i)^m$, $ g_i=(t_c-\tau_i)^m\cos(\omega\ln(t_c-\tau_i))$ and $ h_i=(t_c-\tau_i)^m\sin(\omega\ln(t_c-\tau_i))$. 

The modification from expression  \eqref{eq:lppl} to formula (\ref{eq:lppl_new})
leads to  two very important results. 
\begin{itemize}
\item First, the dimensionality of the nonlinear optimization problem is reduced from 
a 4-dimensional space in~\eqref{eq:S_arg} to a 3-dimensional space in~\eqref{eq:S_arg_new}.
This significantly decreases the complexity of the problem. 
\item Second, and possibly even more important, 
the proposed modification eliminates the quasi-periodicity of the cost function due to subordination of the phase parameter $\phi$ as a part of $C_1$ and $C_2$ to angular log-frequency parameter $\omega$. The existence of multiple minima of the
cost function (as in fig.~\ref{fig:crosssections}) has been the main property
requiring the use of non rigorous metaheuristic searches.
The new formulation does not require such heuristics and rigorous search methods are now sufficient.
\end{itemize}

\begin{figure}[htbp]
  \centering
   \subfigure[$F_1(t_c=\mbox{'06-Nov-2007'}, m, \omega)$]
   { \includegraphics[width=0.48\textwidth]{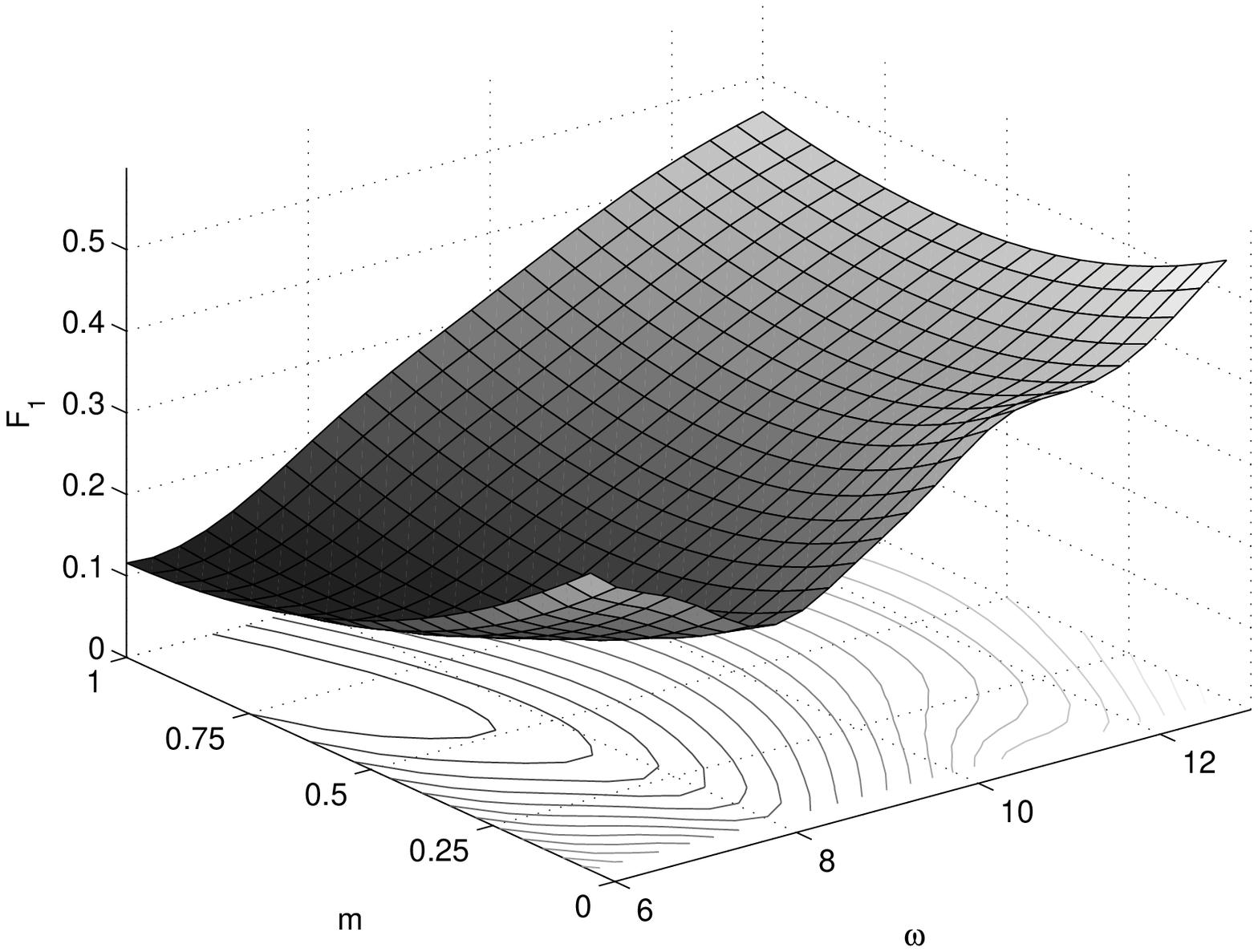}} 
   \subfigure[$F_1(t_c, m=0.7, \omega)$]
   { \includegraphics[width=0.48\textwidth]{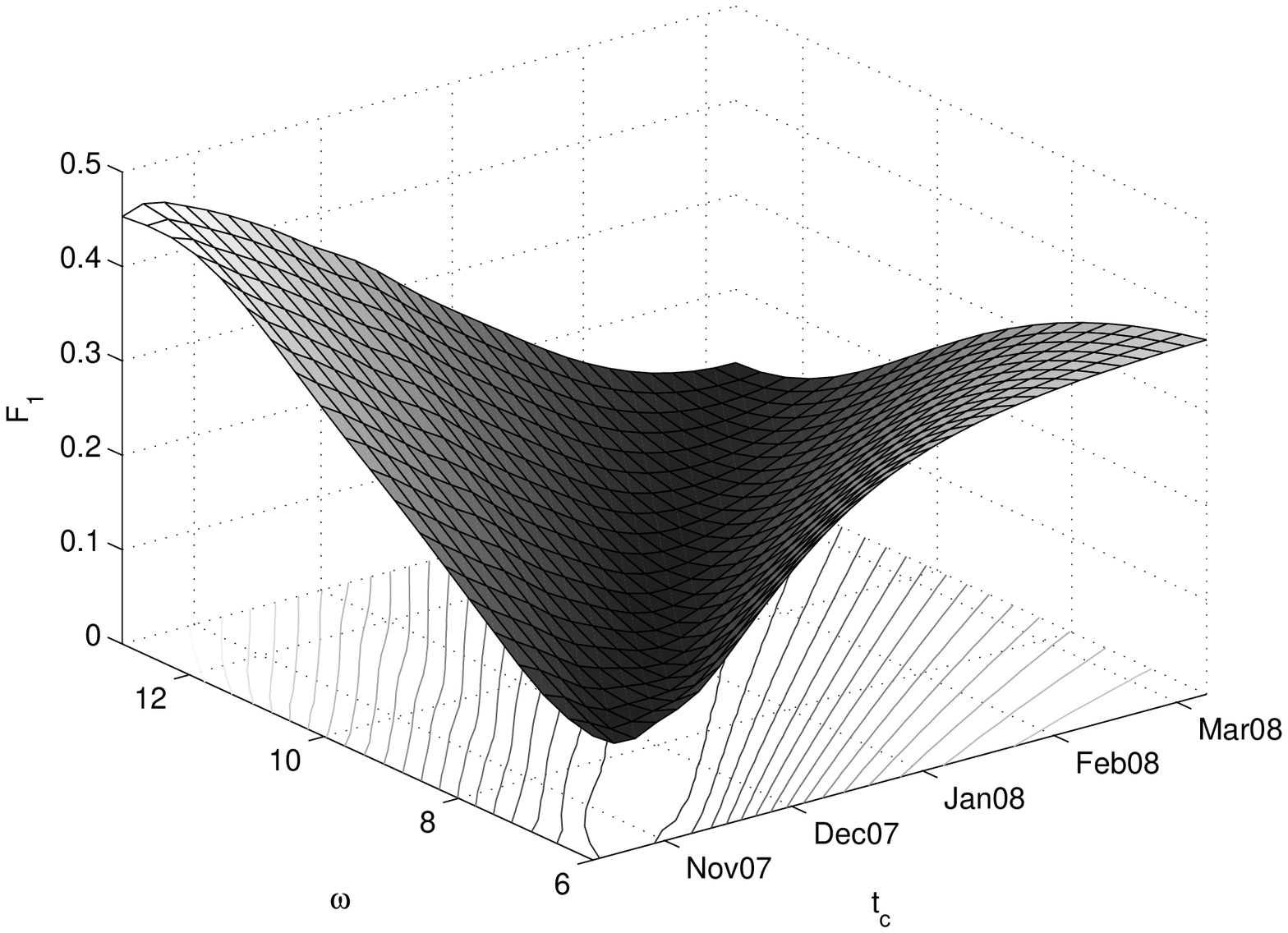}}\\
   \subfigure[$F_1(t_c, m, \omega=7.5)$]
   { \includegraphics[width=0.48\textwidth]{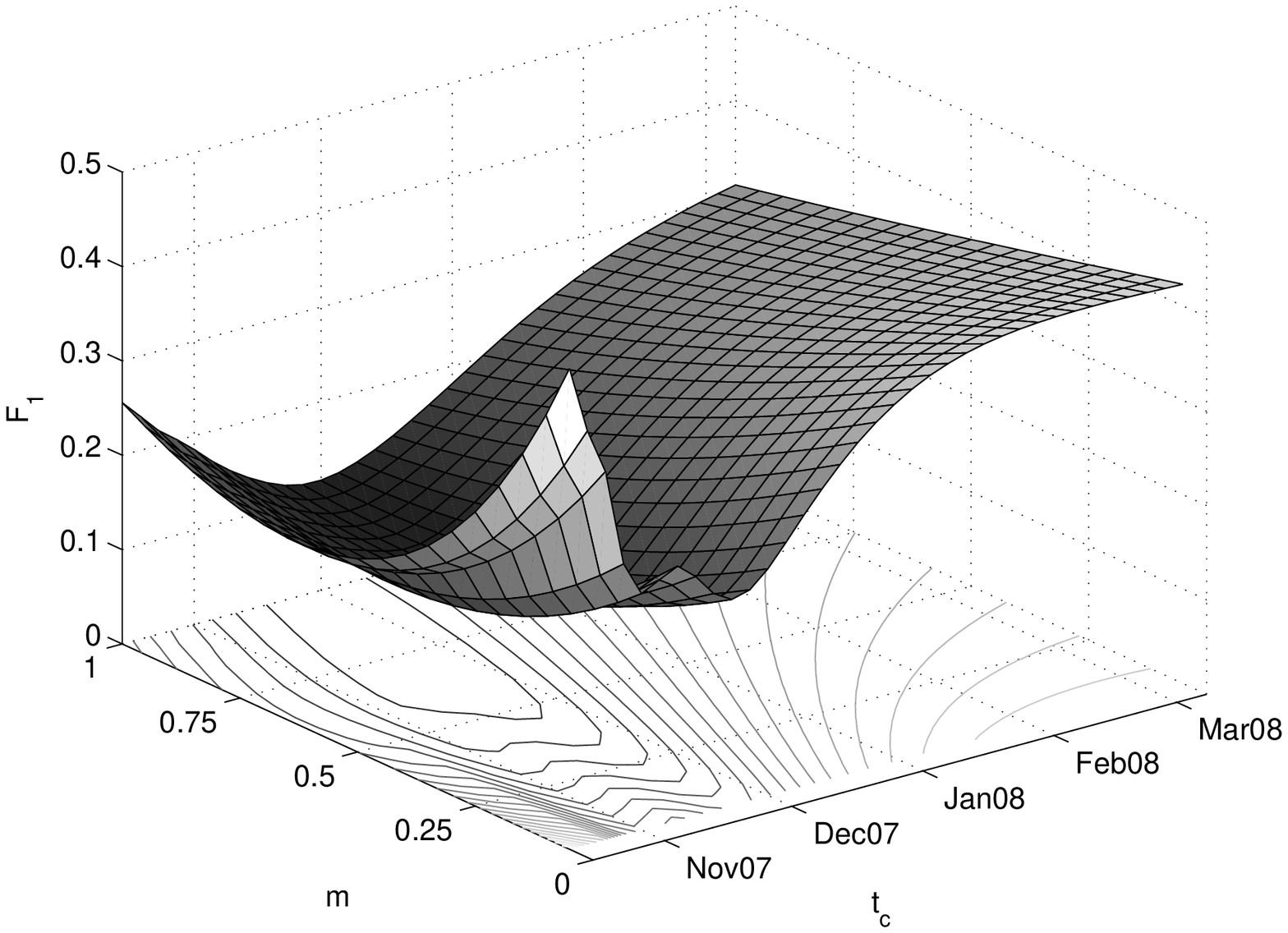}}
   \caption{Different cross-sections of the cost function $F_1$~\eqref{eq:S1_new} for the SSE Composite Index index shown in fig.~\ref{fig:ssec}.} \label{fig:crosssections_new}
\end{figure}

Fig.~\ref{fig:crosssections_new} presents various cross-sections of the cost function $F_1$~\eqref{eq:S1_new} for the same data set that was used in the previous section, namely the SSE Composite Index presented in fig.~\ref{fig:ssec}. We use the parameters that were used for fig.~\ref{fig:crosssections}. One can observe that the cost function now enjoys a very smooth structure with only a few (in the fig.~\ref{fig:crosssections_new} --- only one) minima, which can easily be found using local search methods, such as Levenberg-Marquardt nonlinear least squares algorithm~\cite{Levenberg1944,Marquardt1969} or the Nelder-Mead simplex method~\cite{NelderMead1965}.

\subsection{Decomposing the optimization search by singularizing the critical time $t_c$}

Being the crucial parameter for forecasting the termination of a bubble, the critical time $t_c$ requires special 
care and attention during the fitting procedure. We propose to reformulate the optimization problem~\eqref{eq:S_arg_new} 
by using a similar subordination idea that previously allowed to separate the linear and nonlinear parameters in~\eqref{eq:S_arg_new},\eqref{eq:S1_new} and ~\eqref{eq:S_arg},\eqref{eq:S1}. In this goal and without loss of generality, we rewrite problem~\eqref{eq:S_arg_new} as:
\begin{equation}\label{eq:S_arg2_new}
   \hat t_c=\displaystyle\arg\min_{t_c}\tilde F_2(t_c),
\end{equation}
\begin{equation}\label{eq:S2_new}
    F_2(t_c)=\displaystyle\min_{\omega,m} F_1(t_c,m,\omega),
    \quad 
    \{\hat m(t_c),\hat \omega(t_c)\}=\arg\min_{m,\omega}F_1(t_c,m,\omega)
\end{equation}
where $F_1(t_c, m, \omega)$ is given by~\eqref{eq:S1_new}. The optimization procedure~\eqref{eq:S2_new} operates on the parameter space of the variables $m$ and $\omega$. As seen from fig.~\ref{fig:crosssections_new}a, the cost function $F_1(t_c,m,\omega)$ has a very smooth shape and, in the presented case, only one local minimum. Since, in the general case,
the cost function $F_1(t_c,m, \omega)$ could have more than one minimum within the range of parameters~\eqref{eq:constrains},
we have performed extensive numerical analyses, using the SSE Composite index from July 1999 to May 2011
as the empirical case study. We used a moving window $[t_1,t_2]$ with length of 6 months, scanning the whole range of dates.
In each window, we counted the number of local minima of $F_1(t_c,m,\omega)$ that
fall in the range of parameters~\eqref{eq:constrains} for $t_c$ varying from $t_2+1$ day to $t_2+90$ days. 
The main result is that we never found more than three clearly distinguishable local minima of the function $F_1(t_c,m,\omega)$. 
In other words, the degeneracy in the worst possible cases is very low.
We have found that, when several minima are present, they can be easily determine 
by launching no more than 20 searches with local algorithms started from random points $\{m_0, \omega_0\}$ within the region of $0.1\leq m_0\leq0.9,\quad 6\leq\omega_0\leq13$. We should also stressed that most of the windows 
that are qualified to be in the bubble regime give only one minimum.  The
occurrence of two to three competing minimum has been found to correspond to poor fits of the empirical
price time series by the LPPL function. 

\begin{figure}[htbp]
  \centering
  \includegraphics[width=0.68\textwidth]{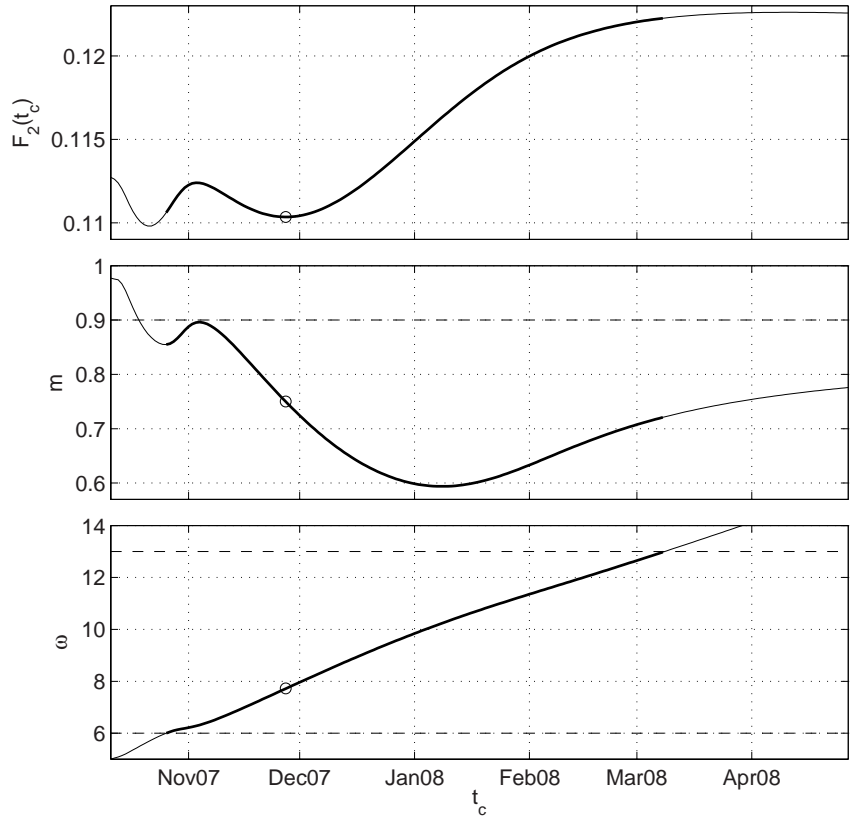}
   \caption{Plot of the cost function $F_2(t_c)$ (top) and
   of the parameters $\hat m(t_c)$ (middle) and $\hat \omega(t_c)$ (bottom) 
   as a function of the critical time $t_c$ (defined as the end of the bubble) for the SSE Composite Index presented in fig.~\ref{fig:ssec}. 
   The circles correspond to the global minimum $\{\hat t_c,\hat m(\hat t_c),\hat \omega(\hat t_c)\}$ of the cost function $F_1(t_c,m,\omega)$. Dashed lines delineate the domains defined by the  constrains~\eqref{eq:constrains}. The thick line represents the values of 
   the parameters where these constrains are met.}  \label{fig:cost_mw}
\end{figure}

The cost function $F_2(t_c)$ of the optimization procedure~\eqref{eq:S_arg2_new} is also very smooth and has only a few minima (see fig.~\ref{fig:cost_mw} for illustration). Again, it is easy to identify these minima 
by using local search algorithms that start from several different initial points  $t_{c0}$. 
Such subordination does not decrease really the computational complexity of the search in~\eqref{eq:S_arg2_new}-\eqref{eq:S2_new} in comparison 
with the 3-dimensional problem~\eqref{eq:S_arg_new}. In general, the required computations are comparable in both approaches. The most important consequence of the reformulation (\ref{eq:S_arg2_new}) with (\ref{eq:S2_new})
is the possibility of studying the quality of the fit of the critical time $t_c$ 
separately and the dependence of the other parameters ($m$ and $\omega$) on the critical time: $\hat m(t_c)$ and $\hat \omega(t_c)$. Fig.~\ref{fig:cost_mw} illustrates the dependence of the cost function $F_2(t_c)$ and of the estimated
parameters $\hat m(t_c)$ and $\hat \omega(t_c)$ as
a function of the critical time $t_c$ for the price time series presented in fig.~\ref{fig:ssec}.

\section{Concluding remarks}
\label{sec:conclusion}


This paper provides an important reformulation of the log-periodic power law (LPPL) equation of the Johansen-Ledoit-Sornette model of financial bubbles which, in its original form, is parametrized by 3 linear and 4 nonlinear parameters. Diagnosing a bubble and forecasting its burst (that usually results in dramatic change of regime, such as large crash or change of the average growth rate) require calibration of the LPPL model and estimation these 7 parameters from the observed price time-series. Though 3 linear parameters could be easily slaved to 4 nonlinear ones, the resulting 4-dimensional search space still keeps a complex structure with quasi-periodicity and multiple minima. The complex structure of the parameter space does not allow one to use  local search algorithms and requires more sophisticated methods, such as taboo search or genetic algorithms. The implementation and the tuning of these methods were the main source of complexity in the calibration of the LPPL model.

The reformulation of the LPPL equation proposed here allows to re-parametrize the model in terms of 4 linear and only 3 nonlinear parameters. This has two very important consequences, namely: (i) reduce of dimensionality of the nonlinear optimization problem from 4-dimensional to 3-dimensional space and (ii) elimination of the quasi-periodicity and of the multiple local minima of the cost function. An empirical case study  performed in this paper using the SSE Composite index from July 1999 to May 2011 shows that, even in the worst cases that were studied, we found no more than 3 competing minima. Most cases that are qualified to be in a bubble regime give only one minimum, and the occurrence of a second or third minimum usually correspond to poor fits of the LPPL function on the empirical price time-series. This simplification allows us to use rigorous local search methods in fitting procedure without introducing any metaheuristics.

As an additional step, we have proposed a subordination procedure that slaves two of the nonlinear parameters, namely the growth rate exponent $m$ and the log-frequency $\omega$, to the critical time $t_c$, that defines the end of the bubble and is considered as the most crucial parameter. This additional subordination provides the following benefits: (i) further decrease of the complexity and (ii) an intuitive representation, that includes as a bonus an explicit dependence of the parameters $m$ and $\omega$ on the critical time $t_c$.

Summarizing, the proposed reformulation of the LPPL model decreased dramatically the complexity of the problem and therefore the effort required for its implementation and use for the diagnosis and forecasting of bubbles in financial markets.

\section*{Acknowledgments}

The authors would like to thank Ryan Woodard, Peter Cauwels and Wanfeng Yan for useful discussions. 

\clearpage

\section*{References}

\begin{thebibliography}{10}
\bibitem{Galbraith1997}
J.~Galbraith, {The Great Crash of 1929}, Mariner Books, 2009.

\bibitem{Sorwoodtokyo}
D.~Sornette, R.~Woodard, Financial bubbles, real estate bubbles, derivative
  bubbles, and the financial and economic crisis, Proceedings of APFA7
  (Applications of Physics in Financial Analysis), ``New Approaches to the
  Analysis of Large-Scale Business and Economic Data,'' Misako Takayasu,
  Tsutomu Watanabe and Hideki Takayasu, eds., Springer (2010)
  (http://arxiv.org/abs/0905.0220).

\bibitem{Sornette_Crash2002}
D.~Sornette, {Why Stock Markets Crash: Critical Events in Complex Financial
  Systems}, Princeton University Press, 2002.

\bibitem{Kindleberger2000}
C.~Kindleberger, {Manias, panics and crashes: a history of financial crises},
  4th Edition, a history of financial crises, John Wiley {\&} Sons Inc, New
  York, 2000.

\bibitem{SornetteJohansen1999Risk}
A.~Johansen, D.~Sornette, {Critical Crashes}, Risk 12~(1) (1999) 91--94.

\bibitem{SornetteJohansen2000}
A.~Johansen, O.~Ledoit, D.~Sornette, {Crashes as Critical Points},
  International Journal of Theoretical and Applied Finance 3~(2) (2000)
  219--255.

\bibitem{SornetteJohansen1999}
A.~Johansen, D.~Sornette, O.~Ledoit, {Predicting Financial Crashes Using
  Discrete Scale Invariance}, Journal of Risk 1~(4) (1999) 5--32.

\bibitem{SornetteJohansen2001}
D.~Sornette, A.~Johansen, {Significance of log-periodic precursors to financial
  crashes}, Quantitative Finance 1~(4) (2001) 452--471.

\bibitem{Jacobsson2009}
E.~Jacobsson, {How to predict crashes in financial markets with the
  Log-Periodic Power Law}, Master's thesis, Department of Mathematical
  Statistics, Stockholm University (2009).

\bibitem{SornetteJohansen2010}
A.~Johansen, D.~Sornette, {Shocks, Crashes and Bubbles in Financial Markets},
  Brussels Economic Review 53~(2) (2010) 201--253.

\bibitem{SornetteLin2009}
L.~Lin, R.~Ren, D.~Sornette, {A Consistent Model of 'Explosive' Financial
  Bubbles with Mean-Reversing Residuals}, Swiss Finance Institute Research
  Paper.

\bibitem{SornetteJohansen2000_geophisics}
Y.~Huang, A.~Johansen, M.~Lee, H.~Saleur, D.~Sornette, {Artifactual
  log-periodicity in finite size data: Relevance for earthquake aftershocks},
  Journal of Geophysical Research-Solid Earth 105 (2000) 25451--25471.

\bibitem{Bothmer2003}
H.~van Bothmer, C.~Meister, {Predicting critical crashes? A new restriction for
  the free variables}, Physica A: Statistical Mechanics and its Applications
  320 (2003) 539--547.

\bibitem{Press2007NumericalRecipes}
W.~H. Press, S.~A. Teukolsky, W.~T. Vetterling, B.~P. Flannery, {Numerical
  Recipes: The Art of Scientific Computing}, 3rd Edition, Cambridge University
  Press, 2007.

\bibitem{Talbi2009Metaheuristics}
E.-G. Talbi, {Metaheuristics: from design to implementation}, Wiley, 2009.

\bibitem{Cvijovicacute1995}
D.~Cvijovicacute, J.~Klinowski, {Taboo Search: An Approach to the Multiple
  Minima Problem}, Science 267~(5198) (1995) 664--666.

\bibitem{SornetteZhouIJF06}
D.~Sornette, W.-X. Zhou, {Predictability of Large Future Changes in major
  financial indices}, International Journal of Forecasting 22 (2006) 153--168.

\bibitem{Levenberg1944}
K.~Levenberg, {A Method for the Solution of Certain Non-Linear Problems in
  Least Squares}, The Quarterly of Applied Mathematics 2 (1944) 164--168.

\bibitem{Marquardt1969}
D.~W. Marquardt, {An Algorithm for Least-Squares Estimation of Nonlinear
  Parameters}, SIAM J. on Applied Mathematics 11~(2) (1962) 431--441.

\bibitem{Gulsen1995}
M.~Gulsen, A.~E. Smith, D.~M. Tate, {A genetic algorithm approach to curve
  fitting}, International Journal of Production Research 33~(7) (1995)
  1911--1923.

\bibitem{NelderMead1965}
J.~Nelder, R.~Mead, {A Simplex Method for Function Minimization}, The Computer
  Journal 7~(4) (1965) 308--313.

\bibitem{Rosser-2008-ACS}
J.~B. Rosser, {Econophysics and economic complexity}, Advances in Complex
  Systems 11 (2008) 745--760.

\bibitem{Lux-2009}
T.~Lux, {Applications of statistical physics in finance and economics}, In:
  Rosser Jr, J.B. (Ed.), Handbook of Complexity Research, Edward Elgar,
  Cheltenham (2009) 213--258.

\bibitem{BreeJoseph2010}
D.~S. Br\'ee, J.~N. Lael, {Fitting the Log Periodic Power Law to financial
  crashes: a critical analysis}, preprint.

\bibitem{BreeChallet2010}
D.~S. Br\'ee, D.~Challet, P.~P. Peirano, {Prediction accuracy and sloppiness of
  log-periodic functions}, preprint.

\bibitem{SornetteZhou2010}
Z.-Q. Jiang, W.-X. Zhou, D.~Sornette, R.~Woodard, K.~Bastiaensen, P.~Cauwels,
  {Bubble diagnosis and prediction of the 2005--2007 and 2008--2009 Chinese
  stock market bubbles}, Journal of Economic Behavior {\&} Organization 74~(3)
  (2010) 149--162.

\end{thebibliography}

\end{document}